\documentstyle[12pt]{article}
\title{Statistical Entropy of Calabi-Yau Black Holes}
\author {Mikhail Z. Iofa\thanks{E-mail address: iofa@theory.npi.msu.su}\,\,
and
Leopoldo A. Pando Zayas\thanks{E-mail address:
leopoldo@grg1.phys.msu.su}\\
Nuclear Physics Institute \\
Moscow State University\\
Moscow 119899, Russia }

\date{}

\def\O{\Omega}

\def\p{\phi}

\def\r{\rho}
\def\g{\gamma}

\def\a{\alpha}
\def\m{\mu}
\def\r{\rho}
\def\s{\sigma}
\begin{document}

\maketitle

\begin{abstract}
We computed the statistical entropy of nonextremal 4D and extremal
5D Calabi-Yau black holes
and found exact agreement with the Bekenstein-Hawking entropy. The
computation is based on the fact that the near-horizon geometry of
equivalent representations contains as a factor the
Ba\~nados-Teitelboim-Zanelli black hole and on subsequent use of
Strominger's proposal generalizing the statistical count of microstates of
the BTZ black hole due to Carlip.
\end{abstract}

\vspace{4cm}
\noindent
PACS numbers: 04.70.Dy, 11.25.Mj \\
Keywords: microscopic entropy, N=2 black holes, $AdS_3$ geometry
\vfill
\noindent NPI-MSU-98-38/492\\
hep-th/9804129

\section{Introduction}
The calculation of the statistical entropy of black holes by reducing the
problem to the counting of microstates of the Ba\~nados-Teitelboim-Zanelli
(BTZ)  black hole \cite{btz} has recently attracted the attention of many
researchers. A fundamental step was taken by Strominger \cite{stro} who
realized  that the
microscopic calculation of the entropy of the BTZ black hole due to Carlip
\cite{carlip} can be performed equally well for any black hole whose
near-horizon geometry contains an asymptotically $AdS_3$ region. This
proposal is a generalization of a result obtained in \cite{sfetsos} where
using the fact that some solutions of type IIA can be related via U-duality
to 4D (5D) black holes and to the product of $BTZ\times S^2(S^3)$
\cite{udual}, a microscopical count of the entropy of some black holes was
performed. Strominger's proposal has already been successfully applied to
provide a microscopical interpretation for the entropy of several black holes
\cite{bir,bl,teo,s4,kaloper,cardoso}. One of the most important features of
this recent approach
is that it does
not rely on supersymmetry and can be applied to any consistent quantum
theory of gravity. This provides the possibility of
microscopically calculating the entropy of not only BPS and near-BPS black
holes, as in the D-brane approach, but of black holes arbitrarily away from
extremality, as shown in \cite{sfetsos}. In the light of this advantage
it is natural to apply the near-horizon
approach to N=2 black holes for which the brane explanation can break down
due to loop corrections as opposed to the N=4 and N=8 black holes for which
supersymmetry protects the counting.

The study of N=2 black holes was pioneered in the works \cite{kallosh}
where some interesting observations were made about the structure of the
moduli and where a way to construct explicit solutions was opened. This
approach provided the basis of many subsequent investigations where extremal BPS
solutions were constructed (see \cite{cybh} and references therein).
Recently \cite{mswv} a rigorous derivation, accounting for tree-level and
loop corrections, of the microscopic entropy of N=2 black holes was given.
This latter description relies on the M-brane content of the black holes.
The M-theory interpretation of these N=2 black holes has also proved useful
in the construction of the nonextremal solutions \cite{kastor,cvetic}.
The nonextremal ansatz proposed in \cite{kastor} is based on the
fact that N=2 extremal black holes are solutions of M-theory
compactification on $CY_3\times S^1$ and are microscopically represented as
three types of M5-branes wrapping four-cycles in $CY_3$ and $S^1$. This suggested that
for the nonextremal solution an analogy with nonextremal black holes of
toroidally compactified M-theory \cite{ts1} could also work. Still, the boost
parameters characterizing the nonextremal solutions must be subject to
certain constraints and explicit solutions were found only for a very
restricted class of N=2 string vacua. A refinement of this construction was
proposed in \cite{cvetic} where it was shown that the ansatz goes through
the equations of motion in the near-horizon region and is valid only to
near-BPS saturated black holes.  This is a sufficient condition to apply
Strominger's proposal which concentrates on the near-horizon region.

In this paper we apply Strominger's  proposal to provide a microscopic
interpretation of the Bekenstein-Hawking entropy of nonextremal 4D
\cite{kastor, cvetic} and extremal 5D Calabi-Yau black holes \cite{sabra} in
terms of the entropy of the BTZ black hole that enters as a factor in the
near-horizon geometry of some of the representations of the considered black
holes. Exact agreement is found in both cases.

\section{Calabi-Yau Black Holes}
\subsection{Nonextremal  4D solution}

The N=2 supergravity action in D=4 includes in addition to the graviton
multiplet, $n_v$ vector multiplets and $n_h$ hypermultiplets. The
hypermultiplet fields can be consistently taken to be constant leaving one
with the following action

\begin{equation}
S = \frac{1}{32\pi G_4} \int \sqrt{-g} \, d^4x \Big[R - 2 g_{A \bar B}
\partial^{\mu} z^A \partial_{\mu} \bar z^{B} - \frac{1}{ 4}
F^I_{\mu\nu} ({^{\star}G_{I}})^{\mu\nu} \Big],
\end{equation}
with the gauge field $G_{I\, \mu\nu}$ given by
\begin{equation}
G_{I\, \mu\nu} = \mbox{Re} {\cal N}_{IJ} F^{J}_{\mu\nu} -
 \mbox{Im} {\cal N}_{IJ} {^{\star}}F^J_{\mu\nu},
\end{equation}
and $I,J = 0,1 .... n_v$.  The
complex scalar fields $z^{A}$ ($A=1..n_{v}$) parametrize a special
K\"ahler manifold with metric $g_{A\bar{B}} = \partial_{A}
\partial_{\bar{B}} K(z,\bar{z})$, where $K(z,\bar{z})$ is the K\"ahler
potential.  Both, the gauge field coupling and  the K\"ahler
potential are expressed in terms of the holomorphic prepotential $F(X)$
\begin{eqnarray}
e^{-K}& = &i (\bar{X}^I F_I - X^I
\bar{F}_{I}) \nonumber \\
{\cal N}_{IJ}& = &\bar{F}_{IJ} + 2i \frac{({\rm Im} F_{IL} X^{L})
({\rm Im} F_{JM} X^{M})} {{\rm Im} F_{MN} X^{M} X^{N}}\ ,
\end{eqnarray}
with $F_{I} = \frac{\partial F(X)}{ \partial X^{I}}$ and $F_{MN} =
\frac{\partial^{2} F(X)}{\partial X^{M} \partial X^{N}}$. The scalar
fields $z^{A}$ are defined by
\begin{equation}
z^{A} = \frac{X^A}{X^0}.
\end{equation}
Throughout the paper we will consider the following prepotential
\begin{equation}
F(X)=\frac{d_{ABC}X^AX^BX^C}{X^0},
\end{equation}
where $d_{ABC}$ are the topological intersection numbers of the Calabi-Yau
manifold. The ansatz discussed in \cite{kastor,cvetic} is the following

\begin{eqnarray}
\label{4d}
ds^2&=&-e^{-2U}fdt^2+e^{2U}\left(f^{-1}dr^2+r^2d\Omega^2\right) \nonumber \\
e^{2U}&=&\sqrt{H_0d_{ABC}H^AH^BH^C} \nonumber \\
 f&=&1-\frac{\mu}{r} \nonumber \\
z^A&=&iH^AH_0e^{-2U}, \quad
H^A=h^A\left(1+\frac{\mu}{r}\sinh^2\gamma_A\right) \nonumber \\
A^0_t&=&\frac{r\tilde H_0'}{h_0 H_0}\quad,\quad
A^C_\varphi=r^2\cos\theta\
\tilde H^{C^\prime} \nonumber \\
\tilde H^A&=&h^A\left(1+\frac{\mu}{r}\cosh\gamma_A\sinh\gamma_A\right) \nonumber\\
H_0&=&h_0\left(1+\frac{\mu}{r}\sinh^2\gamma_0\right)\quad ,\quad
\tilde H_0 =h_0\left(1+\frac{\mu}{r}\cosh\gamma_0\sinh\gamma_0\right)
\end{eqnarray}
where prime denotes derivation respect to $r$. The nonzero components of the gauge field
strengths are
\begin{equation}
F^0_{tr}=\frac{\tilde H_0'}{H_0^2}\quad,\quad F^A_{\varphi\theta}=
r^2\sin\theta\tilde H^{A^\prime}.
\end{equation}

As was already said, this ansatz is restricted to the solution of certain
conditions on ${\cal N_{AB}}$ that influence the values of $\gamma's$
\cite{kastor}. In the approach of \cite{cvetic} the technical restriction on
${\cal N_{AB}}$ is traded to restricting the solution to the near-horizon
region and the final form of the solution describes a near-extremal
solution. The Bekenstein-Hawking entropy of this 4D solution is

\begin{equation}
\label{en4d}
S=\frac{\pi\mu^2}{G_4}\sqrt{h_0\cosh^2\g_0
d_{ABC}h^A\cosh^2\g_Ah^B\cosh^2\g_Bh^C\cosh^2\g_C},
\end{equation}
and the asymptotic flatness condition is $h_0d_{ABC}h^Ah^Bh^C=1$.

\subsection{Extremal 5D solution}
The  N=2 D=5 supersymmetric Lagrangian describing the coupling of vector
multiplets to supergravity is determined by one function which is given by
the intersection form on a $CY_3$:
\begin{equation}
{\cal V} = d_{IJK} X^I X^J X^K.
\end{equation}
The bosonic action is \footnote{For the complete action, including the
fermionic part, see for example \cite{ka}}
\begin{equation}
e^{-1} {\cal L} = -{1\over 2} R - {1\over 4} G_{IJ} F_{\mu\nu} {}^I
F^{\mu\nu J}-{1\over 2} g_{ij} \partial_{\mu} \phi^i \partial^\mu \phi^j
+{e^{-1}\over 8} \epsilon^{\mu\nu\rho\sigma\lambda} d_{IJK}
F_{\mu\nu}^IF_{\rho\sigma}^JA_\lambda^K
\end{equation}
where $R$ is the scalar curvature, $F_{\mu\nu}^I=2\partial_{[\mu}A_{\nu]}^I$
is the Maxwell field-strength
tensor and $e=\sqrt{-g}$ is the determinant of the F\"unfbein
$e_\mu^{\underline \nu}$. The fields $X^I= X^I(\phi)$ are the
special coordinates satisfying
\begin{equation}
X^I X_I=1 , \qquad d_{IJK} X^I X^J X^K =1
\end{equation}
where, $X_I$, the dual coordinate is defined by,
\begin{equation}
X_I=d_{IJK}X^JX^K.
\end{equation}

The moduli-dependent gauge coupling metric is related to the prepotential via
the relation

\begin{equation}
G_{IJ} = -{1\over 2}{\partial\over \partial X^I}
{\partial\over\partial X^J}(\ln {\cal V})|_{{\cal V} =1}
\end{equation}
The metric $g_{ij}$ is given by
\begin{equation}
g_{ij}=  G_{IJ} \partial_{i}X^I\partial_{j}X^J|_{{\cal V} =1};\qquad
(\partial_i \equiv {\partial \over \partial\phi^i})
\end{equation}

Here we will consider the static spherically symmetric BPS black hole
solution of N=2 supergravity in D=5 \cite{sabra} all other solutions are
particular cases of this one. In \cite{sabra} an electrically charged BPS
solutions of supergravity coupled to an arbitrary number of vector multiplets
was constructed.  The metric is of the form
\begin{equation}
ds^2=-e^{-4U}dt^2+e^{2U}(dr^2+d\Omega_3^2).
\end{equation}
Again our main interest is in the form of the solution in the near-horizon
region.  Here the general principle established in \cite{kallosh} that allows
one to compute the area of the horizons of N=2 extremal black holes as an
extremum of the central charge is enough to obtain the near-horizon geometry
of the solution. Namely
\begin{equation}
e^{2U}|_{hor}=\frac{1}{3}X^AH_A|_{hor}=\frac{1}{3}X^A|_{hor}\frac{q_A}{r^2}
=\frac{Z_{0}}{3r^2},
\end{equation}
here $Z=q_AX^A$ is the central charge of the superalgebra and $Z_{0}$ its
value at the horizon. This way, the near-horizon geometry is
\begin{equation}
\label{5dmet}
ds^2=-(1+\frac{Z_0}{3r^2})^{-2}dt^2+(1+\frac{Z_0}{3r^2})(dr^2+r^2d\Omega_3^2)
\end{equation}
If the value of the moduli at the horizon can be consistently kept fixed
throughout the spacetime, as is the case in some 4D solutions, then the
obtained metric is the double-extreme black hole solution \cite{ka}.
The Bekenstein-Hawking entropy is
\begin{equation}
\label{entro5d}
S=\frac{\pi^2}{2G_5}\left(\frac{Z_{0}}{3}\right)^{\frac{3}{2}}
\end{equation}

\section{Statistical Entropy}
To relate the Bekenstein-Hawking entropy of the black holes reviewed in the
previous section to the counting based in the BTZ black hole entropy it is
necessary to show that in an equivalent representation the
near-horizon geometries contain a BTZ black hole as a factor. For the
nonextremal 4D and extremal 5D black hole we consider here the corresponding
representations are as in \cite{sfetsos}, for the 4D black hole we find a
5D representation whose near horizon region takes the form
$BTZ\times S^2$ and for the 5D black hole  $BTZ\times S^3$.

\subsection{Statistical entropy of 4D black holes}

To relate the 4D Calabi-Yau black hole of the previous section to the BTZ it
is convenient to start with the following 5D metric
\begin{eqnarray}
\label{r40}
ds_5^2&=&\left(h_0d_{ABC}H^AH^BH^C\right)^{-1}(-dt^2+dy^2
+\frac{\mu}{r}(\cosh\g_0dt+\sinh\g_0dy)^2 ) \nonumber \\
&+&\left(h_0d_{ABC}H^AH^BH^C\right)^{2/3}(\frac{dr^2}{1-\frac{\mu}{r}}+r^2d\O_2^2)
\end{eqnarray}

Upon compactification over the compact direction  $y$ one obtains the metric
of the nonextremal ansatz (\ref{4d}). This metric is the direct
generalization to the Calabi-Yau case of the toroidal compactification of
three M5-branes that intersect orthogonally over a common string that carries
momentum \cite{ts1}. As a solution of M-theory this metric must have an 11D
lifting, a nice fact noted in \cite{bl} is that the 6D part of the 11D
M-theory solution decouples and then one can consider the 5D part as a
solution of compactification on either a $T^6$ or a $CY_3$. To see that this
is indeed true it suffices to check that upon compactification over the 6D
part (following the notation of \cite{ts1}) it amounts to
$(F_1F_2F_3)^{-\frac{2}{3}6}(F_2F_3)^2(F_1F_3)^2(F_1F_2)^2=1$ as a factor of
the 5D metric. The near-horizon region is $r\to 0$ or more exactly
\begin{equation}
\frac{ \mu h_\a\sinh^2\g_\a }{r} >>1 \,
\end{equation}
for any $\a=0,A$. In this region, defining
\begin{equation}
\label{ld}
l=2(h_0d_{ABC}p^Ap^Bp^C)^{1/3}
\end{equation}
with $p^A=\mu h^A\sinh^2\g_A$, we have
\begin{equation}
ds_5^2=\frac{2r}{l}\left(-dt^2+\frac{\m}{r}(\cosh\g_0dt+\sinh\g_0dy)^2\right)
+\frac{l^2}{4r^2(1-\frac{\mu}{r})}dr^2+\frac{l^2}{4}d\O_2^2,
\end{equation}
where it is explicitly shown that the near-horizon region includes a factor
$S^2$. To check that the remaining three-dimensional part is a BTZ black hole,
it is useful to
make the following change of variables which follows \cite{sfetsos} and
\cite{bl}
\begin{equation}
\tau=\frac{l}{R}t, \qquad \p=\frac{1}{R}y \qquad
\rho^2=\frac{2R^2}{l}(r+\mu\sinh^2\g_0),
\end{equation}
where $R$ is the compactification radius of $y$. In these variables the 5D
metric is:
\begin{eqnarray}
ds_5^2&=& ds_{BTZ}^2+\frac{l^2}{4}d\O_2^2\nonumber \\
ds_{BTZ}&=& -N^2d\tau^2+\rho^2(d\phi+N_{\phi}d\tau)^2 + N^{-2}d\rho^2 \nonumber \\
N^2&=& \frac{\rho^2}{l^2}+\frac{\m^2R^4\sinh^2 2\g_0}{\rho^2l^4}
-\frac{2\m R^2\cosh 2\g_0}{l^3} \nonumber \\
N_{\phi}&=& \frac{\mu R^2\sinh 2\g_0}{\rho^2l^2}
\end{eqnarray}
The geometry of the BTZ black hole is described in \cite{btz} from where we
find that the mass and the angular momentum are
\begin{eqnarray}
\label{para}
M_{BTZ}&=&\frac{2\mu R^2 \cosh 2\g_0}{l^3} \nonumber \\
J_{BTZ}&=&\frac{\m R^2\sinh 2\g_0}{4G_3l^2}.
\end{eqnarray}
To relate the 3D Newton's constant to the 4D one  we can follow for example
\cite{bl,s4}, where the only  ingredients needed are the fact that the Ricci
scalar in $BTZ\times S^2$ decompose as the sum of the Ricci scalar of each
factor and that the 5D action can be written in terms of the 4D Newton's
constant. The final result is
\begin{equation}
\label{newton}
G_3=G_4\frac{2R}{l^2}
\end{equation}
To provide a statistical explanation for the entropy of black holes one
always needs a central charge. The relevant central charge for the BTZ black
hole was initially found in \cite{hen} where it was obtained as the central
charge of the Virasoro algebra of the conformal field theory generated by
the algebra of diffeomophisms of asymptotically $AdS_3$ spaces. As was
already pointed in \cite{s4}, to move from a $BTZ$ geometry to an
asymptotically $AdS_3$ one needs to go to the region of large $r$ and
therefore moves away from the initial near-horizon  geometry. A more
appropriate approach is the one of \cite{ban} where the same central charge is
obtained from the algebra of global charges and it can be found for any value of
the radius. In either approach the central charge is
\begin{equation}
c=\frac{3l}{2G_3}.
\end{equation}
The zero modes of the Virasoro generators are related to the mass and angular
momentum of the BTZ black hole as (see second reference in \cite{btz})
\begin{eqnarray}
M_{BTZ}&=& \frac{8G_3}{l}(L_0+\bar{L}_0)\nonumber \\
J_{BTZ}&=&L_0-\bar{L}_0.
\end{eqnarray}
Now using Cardy's formula for the degeneration of states in a conformal field
theory of given central charge an oscillator levels one finds
\begin{eqnarray}
\label{cardy}
S&=&2\pi\left(\sqrt{\frac{cn_R}{6}}+\sqrt{\frac{cn_L}{6}}\right)\nonumber \\
&=&\frac{\pi}{4G_3}\left(\sqrt{l(lM_{BTZ}+8G_3J_{BTZ})}+
\sqrt{l(lM_{BTZ}-8G_3J_{BTZ})}\right),
\end{eqnarray}
Inserting (\ref{para}) and (\ref{newton}) we obtain the following entropy
\begin{equation}
S=\frac{\pi\m^2}{G_4}\sqrt{h_0\cosh^2\g_0d_{ABC}h^A\sinh^2\g_A
h^B\sinh^2\g_B h^C\sinh^2\g_C}
\end{equation}
which coincides with the geometrical entropy of the nonextremal 4D
Calabi-Yau black hole (\ref{en4d}) for any value of $\g_0$ and for large
values of $\g_I$ which means, in physical terms, for any electric charge and
for large magnetic charges. In the brane picture this is the so called dilute
gas regime $\g_I >> 1$ \cite{mstro}. This way  one  finds that in this limit
the Bekenstein-Hawking entropy of this black hole can be given a statistical
interpretation in terms of the degrees of freedom associated with the
conformal theory of the BTZ black hole.

\subsection{Statistical entropy of 5D black holes}
The ideology to solve this problem for the extremal N=2 D=5 black holes
is exactly the same as for the 4D black holes and therefore most of the
details are omitted. Before writing a 6D metric which upon compactification
on one compact dimension gives the 5D black hole metric (\ref{5dmet})  let us
consider the candidates obtained from analogy with toroidal compactifications
of M-theory. The 5D black hole in toroidal compactifications of M-theory can
be obtained as compactification of two M-theory configurations: three
orthogonally intersecting M2-branes and a M2-brane orthogonally intersecting
a M5-brane (see \cite{ts1} and \cite{ts2} for a detailed analysis). The
technical reason for which the three M2-brane picture can not be used to
obtain the statistical entropy of the three-charge 5D black hole is that
it is not necessary to introduce a boost between time and a compact
coordinate.  This boost is what determine the $AdS_3$ geometry that we need
to be able to apply Strominger's proposal.  To justify the use of the $2\perp
5$ picture we still need to prove that the 5D part that lifts the 6D metric
to a solution of M-theory decouples as was the case in the 4D case. The
metric of the configuration of a M2 intersecting a M5 with a boost along the
common string is \cite{ts1,ts2}
\begin{eqnarray}
ds_{11}^2&=&T^{2/3}F^{1/3}(-K^{-1}f
dt^2+K\hat{dy}_1^2) + T^{-1/3}F^{-2/3}\left(f^{-1}dr^2+r^2d\O_3^2 \right)
\nonumber \\
&+&T^{2/3}F^{-2/3}dy_2^2+T^{-1/3}F^{1/3}(dy_3^2+dy_4^2+dy_5^2+dy_6^2),
\end{eqnarray}
with $f=1-\m^2/r^2$ and $T^{-1}, F^{-1}$ harmonic functions.  It can be seen
that in the limit we are interested in, all charges equal (\ref{5dmet}), the
5D part described by $(y_2,y_3,y_4,y_5,y_6)$ decouples, in fact this part
decouples for any 5D black hole having two equal charges $(F=T)$. Thus we
conclude that the 6D metric that upon compactification gives the 5D
black hole (\ref{5dmet}) can be lifted as a solution of M-theory. In the
conclusions we will comment of what may be the relation of this 6D solution
to the actual compactification of M-theory on $CY_3$ we are considering.  The
6D metric which upon compactification over the compact direction $y$ gives
the 5D black hole metric (\ref{5dmet})  is
\begin{eqnarray}
ds_6^2&=&T\left(-dt^2+dy^2+\frac{\m^2}{r^2}(\cosh\s dt
+\sinh\s dy)^2\right) \nonumber \\
&+&T^{-1}\left(\frac{dr^2}{f}+r^2d\O_3^2\right).
\end{eqnarray}
It is still necessary to put
\begin{equation}
T^{-1}=1+\frac{Z_0}{3r^2},
\end{equation}
and take the limit: $\m^2\to 0, \s\to\infty$ with $\m^2\sinh^2\s\to Z_0/3$.
Performing the change of variable
\begin{equation}
\r^2=\frac{R^2}{l^2}(r^2+\m^2\sinh^2\s) \qquad \p=\frac{y}{R}
\qquad \tau=\frac{l}{R}t,
\end{equation}
with $R$ the compactification radius and taking $l=(Z_0/3)^{1/2}$ one find
that the 6D metric can be written as
\begin{eqnarray}
ds_5^2&=& ds_{BTZ}^2+\frac{l^2}{4}d\O_3^2\nonumber \\
ds_{BTZ}&=& -N^2d\tau^2+\rho^2(d\phi+N_{\phi}d\tau)^2 + N^{-2}d\rho^2 \nonumber \\
N^2&=& \frac{\rho^2}{l^2}+\frac{\m^4R^4\sinh^2 2\s}{4\rho^2l^6}
-\frac{\m^2 R^2\cosh 2\s}{l^4} \nonumber \\
N_{\phi}&=& \frac{\mu^2 R^2\sinh 2\s}{2\rho^2l^3}.
\end{eqnarray}
The mass and angular momentum of this solution are
\begin{eqnarray}
\label{para2}
M_{BTZ}&=&\frac{\mu^2 R^2 \cosh 2\s}{l^4} \nonumber \\
J_{BTZ}&=&\frac{\m^2 R^2\sinh 2\s}{8G_3l^3}.
\end{eqnarray}
The relation between the Newton's constants is
\begin{equation}
\label{newton2}
\frac{1}{G_3}=\frac{1}{G_5}\frac{\pi l^3}{R}.
\end{equation}
Using (\ref{para2}), (\ref{newton2}) and (\ref{cardy}) the statistical
entropy is
\begin{equation}
S=\frac{\pi^2}{2G_5}l^2\mu\cosh\s,
\end{equation}
which coincides with the entropy (\ref{entro5d} in the limit of large $\s$,
which is indeed needed to obtain the extremal limit.

\section{Conclusions} Here it has been shown that the Bekenstein-Hawking
entropy of Calabi-Yau black holes can be given a statistical interpretation
using Strominger's proposal. For the 4D black hole the nonextremal case was
considered for an ansatz that is more general than the  actual black hole
metric. As shown in \cite{kastor,cvetic} some other conditions most be
included that restrict the ansatz. Exact agreement was found in the "dilute
gas" regime or more precisely for large values of the magnetic charges and
arbitrary values of the electric charge. In the 5D case the extremal solution
was treated using the general form of the near-horizon metric presented in
\cite{sabra} and based on the beautiful results of \cite{kallosh}.  Hopefully
for the nonextremal 5D black holes a counting similar to the one carried here
for the 4D could be performed, still some work is needed to consistently find
the explicit form of the 5D nonextremal Calabi-Yau black holes. In the
present work one step has been made in this direction noting that the M-brane
picture that is consistent with the counting of the microstate selects the
microscopic content of the black hole as a bound state of M2 and M5-branes
and not that of only M2-branes. It could also be conjectured that a
twelve-dimensional theory may be of importance in the construction of
nonextremal 5D black holes since in the counting use was made of a theory of
that has a $S^1$ factor and upon compactification on it yields the 5D black
hole. The 5D black hole in the picture used in this paper seems to be a
solution of a twelve-dimensional theory compactified on a  $CY_3\times S^1$.
Another sensible picture is that of M-theory compactified on a $CY_3$ that
can effectively be described as having a $S^1$ factor.

Since N=2 black holes receive quantum corrections it is worth explaining
to what extend this microscopic counting could include them. Unfortunately
a rigorous statistical explanation of the quantum corrections as
presented in \cite{mswv} does not seem to be available in this picture.
Still, it is worth noting that certain quantum corrections can be included in
the approach used in this paper. Namely those that preserve the polynomial degree of three
of the prepotential. One example of this type of quantum corrections was presented
in \cite{oneloop} and has the form
\begin{equation}
d_{ABC}H^AH^BH^C=H^1H^2H^3+a(H^3)^3.
\end{equation}
The metric for this quantum corrected solution is
\begin{equation}
g_{00}^2=4(h_0+{q_0\over r}) \left((h_1+{p_1\over r})(h_2+{p_2\over r})
(h_3+{p_3\over r})+a(h_3+{p_3\over r})^3\right),
\end{equation}
redefining $l$ in (\ref{ld}) one can consistently include this quantum
correction to the geometric entropy in the statistical description
presented here.

\begin{center}
{\large \bf Acknowledgments}
\end{center}
This work was  partially supported by the RFFR  grant No 98-02-16769. LAPZ
is grateful to N. Pando Girard for reading this paper.

\end{document}